# Crosscorrelation kernel in the Green's function retrieval and time reversal acoustics

Yingcai Zheng

Earth and Planetary Sciences Department, University of California, Santa Cruz, California, 95064, USA.

**Abstract**

Crosscorrelation structures in the Green's function retrieval by crosscorrelating wavefields are revealed using rigorous mathematical theory on integral equations. The previous practice on extracting the Green's function by crosscorrelating the wavefields recorded at two locations produced by the same source and then summing such crosscorrelations over all sources on a boundary is inadequate and has limitations in recovering the low frequency content in the Green's function. To recover the exact Green's function, we need crosscorrelate the wavefields observed at the two locations generated by different sources, respectively. The crosscorrelation structure can be viewed as a matrix in the discrete case or an integral operator in the continuous case. The previous Green's function retrieval method corresponds to the identity matrix multiplying a constant. If the matrix is diagonal, the wavefield crosscorrelation is still within the wavefield due to the same source but different weighting should be applied for different sources. We have derived analytically the crosscorrelation kernels for two important cases, the plane boundary and the circular boundary, and both kernels are symmetrical difference kernels, which correspond to convolutional operations. For high frequency waves or far-field sources, the kernels reduce to the know result which is the Dirac delta distribution (or an identity matrix). For other boundaries of general geometric shapes, numerical schemes like the boundary element method can be used to solve for the kernel matrix.

It has been shown that by crosscorrelating the diffuse wavefield recordings at two locations, one can obtain the Green's function that would be observed at one location as if a source were at the other location[1, 2]. Using crosscorrelation methods to extract full or partial response of the medium represents an active area of research and they have been applied in different fields to obtain new information about the medium such as in ultrasonics [2] and time reverse acoustics [3], surface wave extraction by seismic coda correlation in global seismology [4, 5], time-distance analysis or daylight imaging in helioseismology[6, 7], seismic interferometry/imaging [e.g., 8, 9] and the virtual source method [10, 11] in the exploration seismology, to name a few. For a one-dimensional layered medium, Claerbout [12] rigorously showed how the autocorrelation of the transmission response could yield the point source response of the medium. Lobkis and Weaver [2] gave the field a new momentum by deriving the Green's function retrieval using normal modes for a closed system. Other theoretical derivations for three-dimensional open system include stationary phase methods [13, 14] and the one based on the representation theorem by Wapenaar [15] which unified several crosscorrelation based concepts that are related to the Green's function retrieval but his derivation involved high frequency approximations and in general neglected inhomogeneous waves. In this Letter, we show that this approximation contains important crosscorrelation structure and how we can remove the approximation using the exact integral equation method.

The Green's function retrieval can be represented by an surface integral (**Figure 1**) in the frequency $\omega$ domain using the Rayleigh's reciprocity of the correlational type [e.g., 16, 17]:

$$2i\,\mathrm{Im}\,G(\mathbf{x}_B|\mathbf{x}_A;\omega) = \oiint_{\partial\mathbb{D}} \left[ G(\mathbf{x}'|\mathbf{x}_A;\omega)\frac{\partial \overline{G}(\mathbf{x}'|\mathbf{x}_B;\omega)}{\partial n'} - \overline{G}(\mathbf{x}'|\mathbf{x}_B;\omega)\frac{\partial G(\mathbf{x}'|\mathbf{x}_A;\omega)}{\partial n'} \right] d^2\mathbf{x}' \quad (1)$$

in which $G(\mathbf{x}'|\mathbf{x}_A;\omega)$ and $G(\mathbf{x}'|\mathbf{x}_B;\omega)$ are the Green's functions for the heterogeneous medium in $\mathbb{D}$ which satisfy differential equations

$$\nabla^2 G(\mathbf{x}'|\mathbf{x}_{A,B};\omega) + k^2(\mathbf{x}') G(\mathbf{x}'|\mathbf{x}_{A,B};\omega) = -\delta(\mathbf{x}'-\mathbf{x}_{A,B}), \quad \mathbf{x}',\mathbf{x}_{A,B} \in \mathbb{D}\cup\partial\mathbb{D} \text{ and}$$

the over bar denotes the complex conjugate and $\partial/\partial n'$ is the directional derivative along the outward normal of the boundary $\partial\mathbb{D}$ with respect to position $\mathbf{x}' \in \partial\mathbb{D}$. For simplicity, we neglect the source time function in the formulation. The right hand side of (1) can be interpreted as focusing/backpropagating the wavefield on $\partial\mathbb{D}$ generated by a source at $\mathbf{x}_A$ to the observation point $\mathbf{x}_B$ and by a physical argument using the time reversal acoustics [3] the left hand side of (1) corresponds to $G(\mathbf{x}_B|\mathbf{x}_A;t) - G(\mathbf{x}_B|\mathbf{x}_A;-t)$ in the time domain, which consists of the causal and the anticausal Green's functions. Let $p_A(\mathbf{x}') = G(\mathbf{x}'|\mathbf{x}_A;\omega)$ and $q_A(\mathbf{x}') = \partial G(\mathbf{x}'|\mathbf{x}_A;\omega)/\partial n'$ be the wavefield on $\partial\mathbb{D}$ produced by a source at $\mathbf{x}_A$. Likewise, $p_B(\mathbf{x}') = G(\mathbf{x}'|\mathbf{x}_B;\omega)$ and $q_B(\mathbf{x}') = \partial G(\mathbf{x}'|\mathbf{x}_B;\omega)/\partial n'$ are due to the source at $\mathbf{x}_B$. The backpropagation integral can be written in a compact form

$$2i\,\mathrm{Im}\,G(\mathbf{x}_B|\mathbf{x}_A;\omega) = \langle p_A, q_B \rangle - \langle q_A, p_B \rangle \quad (2)$$

in which $\langle f,g \rangle = \int_{\partial\mathbb{D}} f(\mathbf{x}')\overline{g}(\mathbf{x}') d^2\mathbf{x}'$ is the inner product defined on the boundary. In previous studies [e.g., 16, 18, 19], the integral can be simplified in an open system using the far-field or the high frequency approximation for a spherical $\partial\mathbb{D}$ with a large radius such that $q_{A,B}(\mathbf{x}') \approx i\omega c^{-1}(\mathbf{x}') p_{A,B}(\mathbf{x}')$, $\mathbf{x}' \in \partial \mathbb{D}$. If the wave propagation speed $c(\mathbf{x}')$ is constant, we obtain

$$2i \operatorname{Im} G(\mathbf{x}_B | \mathbf{x}_A; \omega) \approx -2ik \langle p_A, p_B \rangle, \quad k = \omega / c \qquad (3)$$

which is referred as Green's function retrieval from far-field correlations. By the source-receiver reciprocity in the Green's function, $p_A(\mathbf{x}')$ and $p_B(\mathbf{x}')$ can be interpreted as the wavefields recorded at $\mathbf{x}_A$ and $\mathbf{x}_B$, respectively, produced by a source at $\mathbf{x}' \in \partial \mathbb{D}$. Each source contributes equally to the Green's function retrieval on the left hand side. However, if $\partial \mathbb{D}$ is not in the far-field region, can we still relate the backpropagation integral (1) to the crosscorrelation of the wavefield as in (3)?

It is well known in the studies of Huygens principle using single-layer and double-layer potential representations that the surface values $p_A$ and $q_A$ (or $p_B$ and $q_B$) are generally not independent [e.g., 20, 21, 22]. To see this, we consider the scattering problem due to a source at $\mathbf{x}_A \in \mathbb{D}$. In the exterior domain $\mathbb{D}^c$, we apply the boundary integral equation method [e.g., 23, 24, 25]:

$$\sigma(\mathbf{x}'') p_A(\mathbf{x}'') = \oiint_{\partial \mathbb{D}} \int \left[ p_A(\mathbf{x}') \frac{\partial G(\mathbf{x}' | \mathbf{x}''; \omega)}{\partial n'} - q_A(\mathbf{x}') G(\mathbf{x}' | \mathbf{x}''; \omega) \right] d^2\mathbf{x}' \quad \mathbf{x}', \mathbf{x}'' \in \partial \mathbb{D} \qquad (4)$$

in which the continuity of $p_A$ and $q_A$ across the boundary is imposed and the integral is in the sense of Cauchy principal value. If the boundary is a corner point at $\mathbf{x}''$, $\sigma(\mathbf{x}'')$ is related to the solid angle spanned by the corner and $\sigma(\mathbf{x}'') = 0.5$ for a smooth boundary point. Equation (4) can be written in the integral operator form

$$Sq_A(\mathbf{x}'') = \left(K - \frac{1}{2}I\right) p_A(\mathbf{x}''), \quad \mathbf{x}'' \in \partial \mathbb{D}.$$

The exterior Green's function satisfies the differential equation in the exterior domain

$$\nabla'^2 G(\mathbf{x}';\mathbf{x}'') + k^2(\mathbf{x}') G(\mathbf{x}';\mathbf{x}'') = -\delta(\mathbf{x}'-\mathbf{x}''), \quad \mathbf{x}',\mathbf{x}'' \in \mathbb{D}^c \cup \partial \mathbb{D}$$

and possibly the Sommerfeld radiation condition at the infinity

$$\lim_{r \to \infty} r\left(\frac{\partial}{\partial r} - ik\right) G(\mathbf{x}';\mathbf{x}'') = 0, \quad r = |\mathbf{x}'-\mathbf{x}''|$$

if the exterior medium is unbounded. The null space of the operator $S$ is not empty if $\omega$ is one of the eigenfrequencies $\{\omega_k, k \in \mathbb{N}\}$ of the interior problem with the homogeneous Dirichlet boundary condition [20]. Under this case, the inverse $S^{-1}$ of $S$ does not exist and $q_A$ cannot be determined uniquely from $p_A$. Because the backpropagation integral is a general expression for all $\omega$, we can restrict our discussion to the case $\omega \neq \omega_k$. However, if the exterior domain is unbounded and the Sommerfeld radiation condition is ensured at the infinity, the boundary value $q_A$ is uniquely determined by $p_A$ [21]. Once the inverse $S^{-1}$ of $S$ exists, we have

$$q_A = S^{-1}\left(K - \frac{1}{2}I\right) p_A = C p_A \tag{5}$$

in which $I$ is the identity operator. The same reasoning can be applied to the source at $\mathbf{x}_B$:

$$q_B = C p_B. \tag{6}$$

Substituting (5) and (6) into (2) we obtain

$$2i\,\mathrm{Im}\,G(\mathbf{x}_B \mid \mathbf{x}_A;\omega) = \langle p_A, C p_B \rangle - \langle C p_A, p_B \rangle = \langle p_A, (C - C^*) p_B \rangle = \langle p_A, W p_B \rangle$$

in which $C^*$ is the adjoint of $C$. If we discretize the boundary, $C$ will be a square matrix and $C^*$ is the conjugate transpose of $C$. The inverse of the matrix $S$ can be studied readily from its eigenvalues. We take the interpretation that $p_A(\mathbf{x}')$ and $p_B(\mathbf{x}')$ are wavefields at $\mathbf{x}_A$ and $\mathbf{x}_B$, respectively, due to a boundary source at $\mathbf{x}'$. Clearly, if $W = \{w_{ij}\}$ is a diagonal matrix, then the backpropagation is a weighted correlation, $\sum_i w_{ii} p_A^i \bar{q}_B^i$ in which $i$ is the index of the source. However, if $W$ is a full matrix, the backpropagation integral becomes $\sum_i \sum_j w_{ij} p_A^i \bar{q}_B^j$ and $i$ and $j$ are the source indices. In this case, the wavefield $p_A^i$ at $\mathbf{x}_A$ due to the *i*th source on the boundary is crosscorrelated with the wavefield $p_B^j$ at $\mathbf{x}_B$ due to the *j*th source, which is in sharp contrast with the previous thought that only wavefields at two locations from the same source are crosscorrelated. In general, the matrix $W$ depends on the exterior medium and the boundary geometry. Note that if the exterior medium is unbounded and homogeneous, $W$ only depends on the geometry of the boundary $\partial D$ and the wavenumber $k$ in the exterior medium.

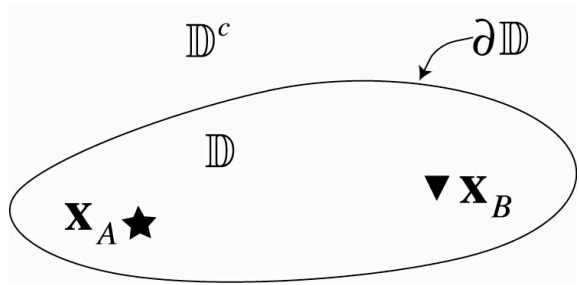

Figure 1. Heterogeneous medium bounded by a surface $\partial \mathbb{D}$.

In what follows, we will show two important boundaries frequently appeared in applications: an infinite plane boundary and a cylindrical boundary. Both problems are formulated in the two-dimensional space and both crosscorrelation kernels are symmetrical difference kernels whose actions on a function represent convolution operations.

Example 1: In the two-dimensional space, let the boundary $\partial \mathbb{D}$ be an infinitely large plane. In this case, $K = 0$. The integral operator kernel is

$$S = \frac{i}{4} H_0(k|x|)$$

in which $H_0$ is the Green's function in the two-dimensional space. Because the kernel $S$ is a symmetrical difference kernel whose action on a function is a convolution, the Fourier transform in the distributional sense can be applied to compute its inverse and the kernel reads:

$$W(x) = ikJ_1(k|x|)/|x|$$

where $J_1$ is the first order Bessel function of the first kind. Figure 2 plots one such a kernel.

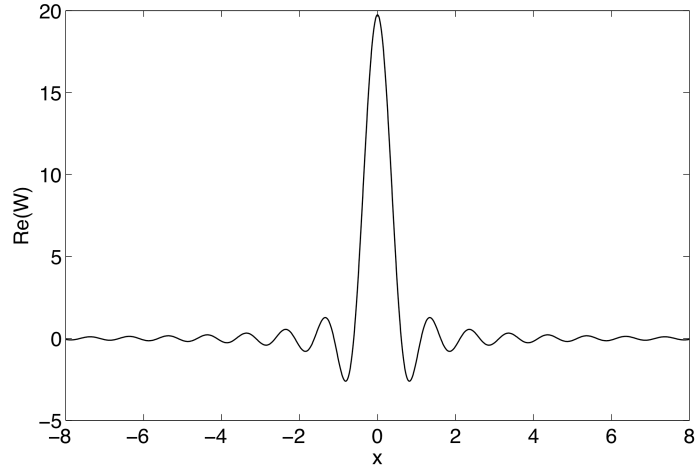

**Figure 2 Crosscorrelation kernel for the plane surface case. The wavelength $\lambda=1$.**

Example 2: If the surface is a cylinder with radius $R$ and the exterior medium is unbounded and homogeneous and has a wavenumber $k$. The integral equation in the exterior domain reads

$$\frac{1}{2}u(\theta) = -\frac{i}{4}\int_0^{2\pi} q(\theta') H_0\left(2kR\sin\frac{\theta-\theta'}{2}\right) R\,d\theta' - \frac{ik}{4}\int_0^{2\pi} u(\theta') H_1\left(2kR\sin\frac{\theta-\theta'}{2}\right)\sin\left(\frac{\theta-\theta'}{2}\right) R\,d\theta'$$

in which polar angles $\theta$ and $\theta'$ denote positions on the boundary. All functions in the above integral are $2\pi$-periodic. Therefore, we can use Fourier series to obtain the crosscorrelation kernel in the discrete angular wavenumber domain

$$W_n = \frac{4i}{\pi R}\left|H_n(kR)\right|^{-2}, \quad n=0,\pm 1,\pm 2,\cdots$$

in which $W_n$ is the Fourier coefficient for the expansion function $e^{in\theta}$ and $H_n$ the first-kind Hankel function of order $n$. For the far-field case, $kR \gg 1$, our crosscorrelation kernel reduces to the familiar one, $W(\theta) = 2ik\delta(\theta)$, where $\delta(\theta)$ is the Dirac delta function. Hence we have rigorously justified the validity of using the far-field approximation in the Green's function retrieval by the integral equation method in equation (3). Let $R = 1$ and we compute the crosscorrelation kernels for several wavenumbers (Figure 3).

To summarize and remark further on the general properties of the crosscorrelation kernels, we consider the situation in which deterministic sources are placed on the boundary and the two receivers are enclosed by the boundary. In this Letter, we have shown that in order to exactly retrieve the Green's function we need correlate wavefields not only between receivers for the same source but also between different sources. If there are $N_S$ sources and $N_R$ receivers and the crosscorrelation kernel is a full matrix, we need calculate $\binom{N_S N_R}{2}$ crosscorrelations rather than $N_S \binom{N_R}{2}$ in the previous practice where $\binom{N}{2}$ is the binomial coefficient. For both the plane and the cylindrical boundaries, the crosscorrelation kernels approach to the Dirac delta distribution if we increase the frequency $\omega$. Numerical examples have been conducted for the cylindrical boundary and corroborated our theoretical prediction that the accuracy of the retrieved Green's function increases for large $kR$ if only using the delta kernel [18]. Our exact theory based on the integral equation method indicates that crosscorrelating wavefields at two locations from the same source corresponds to the high frequency approximation. However, in order to accurately recover the low frequency content in the Green's function, we need crosscorrelate the wavefields from different sources and in the meantime the crosscorrelation kernel should be applied and this has important implications in applications such as waveform tomography using retrieved Green's functions and the normal modes of the Earth which uses extremely long-period seismic waves. At the high frequency, the crosscorrelation kernel for the cylindrical boundary resembles that of the plane boundary case (Figure 3c & Figure2). This indicates that for a segment on the boundary, if the local radius of curvature is large compared to the wavelength, the crosscorrelaiton kernel of the plane boundary may be applied locally. Finally, we note that if the sources on the boundary are uncorrelated noise sources, the crosscorrelation kernel should still be used in the Green's function retrieval.

Acknowlegements: I thank Ru-shan Wu, Xiao-bi Xie, Yaofeng He and Rui Yan for verifying the idea and checking the laborious derivation. I also thank Professors Thorne Lay, Haruo Sato and Mike Fehler for reading the manuscript. I thank Professor Harold Widom for help on the integral equation with Hankel kernel and its inverse. This work is partially supported by NSF grant award no. EAR-0838359 and by the Wavelet Transform on Propagation and Imaging (WTOPI) Consortium/UCSC. Facility support from the W. Keck Foundation is appreciated.

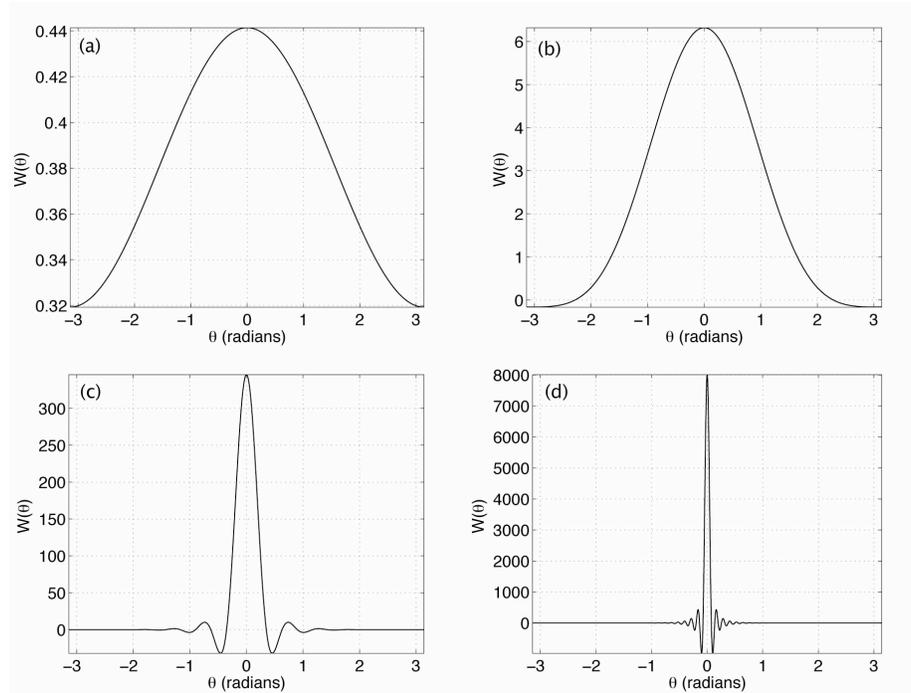

**Figure 3. Crosscorrelation kernels for a cylinder boundary with radius *R*=1 for different wavenumbers (a) *k*=0.1; (b) *k*=1; (c) *k*=10 and (d) *k*=50. The kernels are obtained by inverse Fourier transform of Im$W_n$.**